\documentclass[aps,prb,reprint,amsmath,amssymb,superscriptaddress,showpacs,floatfix,10pt]{revtex4-1}

\usepackage{graphicx}
\usepackage{dcolumn}
\usepackage{bm}
\usepackage{hyperref}
\usepackage[usenames, dvipsnames]{color}

\begin{document}

\title{Boundary-induced spin density waves in linear Heisenberg
antiferromagnetic spin chains with $\mathbf{S \ge 1}$}

\author{Dayasindhu Dey}
\email{dayasindhu.dey@bose.res.in}
\affiliation{S. N. Bose National Centre for Basic Sciences, Block - JD, Sector - III, Salt Lake, Kolkata - 700098, India}

\author{Manoranjan Kumar}
\email{manoranjan.kumar@bose.res.in}
\affiliation{S. N. Bose National Centre for Basic Sciences, Block - JD, Sector - III, Salt Lake, Kolkata - 700098, India}

\author{Zolt\'an G. Soos}
\email{soos@princeton.edu}
\affiliation{Department of Chemistry, Princeton University, Princeton, New Jersey 08544, USA}

\date{\today}

\begin{abstract}
Linear Heisenberg antiferromagnets (HAFs) are chains of spin-$S$ 
sites with isotropic exchange $J$ between neighbors. Open and 
periodic boundary conditions return the same ground state energy 
in the thermodynamic limit, but not the same spin $S_G$ when 
$S \ge 1$. The ground state of open chains of $N$ spins has 
$S_G = 0$ or $S$, respectively, for even or odd $N$. Density matrix 
renormalization group (DMRG) calculations with different algorithms 
for even and odd $N$ are presented up to $N = 500$ for the energy and spin 
densities $\rho(r,N)$ of edge states in HAFs with $S = 1$, 3/2 and 2. 
The edge states are boundary-induced spin density waves 
(BI-SDWs) with $\rho(r,N) \propto (-1)^{r-1}$ for $r=1,2,\ldots N$. 
The SDWs are in phase when $N$ is odd, out of phase when $N$ is even, 
and have finite excitation 
energy $\Gamma(N)$ that decreases exponentially with $N$ for integer 
$S$ and faster than $1 / N$ for half integer $S$. The spin densities 
and excitation energy are quantitatively modeled for integer $S$ 
chains longer than $5 \xi$ spins by two parameters, the correlation 
length $\xi$ and the SDW amplitude, with $\xi = 6.048$ for $S = 1$ 
and 49.0 for $S = 2$. The BI-SDWs of $S = 3/2$ chains are not localized 
and are qualitatively different for even and odd $N$. Exchange between 
the ends for odd $N$ is mediated by a delocalized effective spin in the 
middle that increases $|\Gamma(N)|$ and weakens the size dependence. 
The nonlinear sigma model (NL$\sigma$M) has been applied the HAFs, 
primarily to $S = 1$ with even $N$, to discuss spin densities and exchange 
between localized states at the ends as $\Gamma(N) \propto (-1)^N \exp(-N/\xi)$.
$S = 1$ chains with odd $N$ are fully consistent with the NL$\sigma$M; 
$S = 2$ chains have two gaps $\Gamma(N)$ with the same $\xi$ as 
predicted whose ratio is 3.45 rather than 3; the NL$\sigma$M is more 
approximate for $S = 3/2$ chains with even $N$ and is modified for 
exchange between ends for odd $N$.
\end{abstract}

\pacs{75.10.Pq,75.10.Kt,75.30.Fv,75.40.Mg}

\maketitle

\section{\label{sec:intro}Introduction}
The Hilbert space of a system of $N$ spins $S$ has dimension 
$(2S + 1)^N$. The total spin $S_T \le N S$ and its $z$ components 
are conserved for isotropic (Heisenberg) exchange interactions 
between spins. The simplest case is a chain with equal exchange $J$ 
between nearest neighbors. A great many theoretical and experimental 
studies have been performed on the linear Heisenberg antiferromagnet 
(HAF), Eq. \ref{eq:ham} below, with $S = 1/2$ and $J > 0$. There are 
multiple reasons why. First, there are good physical realizations of 
spin-1/2 chains in inorganic crystals with localized spins on metal 
ions and in organic crystals based on one-dimensional (1D) stacks of 
radical ions. Second, the Hilbert space is smallest for $S = 1/2$ for 
any choice of exchange interactions, small enough to access the full 
spectrum and thermal physics for comparison with experiment. Third, 
long ago Bethe and Hulthen obtained the exact ground 
state~\cite{bethe31,*hulthen38} of the infinite chain with 
antiferromagnetic exchange between nearest neighbors, a prototypical 
gapless many-body system with quasi-long-range order.

HAFs and related chains with $S \ge 1$ came to the fore with Haldane's 
conjecture based on field theory that integer $S$ chains are 
gapped.~\cite{haldane82} Shortly thereafter, White introduced the density 
matrix renormalization group (DMRG) method that made possible accurate 
numerical calculation of the ground state properties of $S \ge 1$ 
chains.~\cite{white-prl92,*white-prb93} The thermodynamic limit of spin 
chains with exchange interactions leads to quantum phase diagrams 
with many interesting correlated phases. According to the valence bond 
solid (VBS) analysis,~\cite{affleck88} integer $S$ chains have localized 
edge states with spin $s = S/2$. DMRG studies of finite chains have 
confirmed edge states in both integer~\cite{white-huse-prb93,schollwock96} 
$S$ and half integer~\cite{qin95,fath2006} $S$ chains. Machens 
et al.,~\cite{machens2013} have recently discussed short $S \ge 1$ HAFs 
with comparable energies for bulk excitations and edge states. They 
summarize previous studies such as the relation of $S \ge 1$ HAFs to the 
nonlinear $\sigma$ model (NL$\sigma$M), its application to edge states, 
the VBS model and its valence bond diagrams. Qin et al.,~\cite{qin95} 
applied DMRG to HAFs up to 100 spins to discuss the energies of edge 
states and to distinguish between chains of integer and half integer $S$. 
DMRG is quantitative for $S = 1$ HAFs of $N \le 100$ spins with 
correlation length $\xi \sim 6$ and large Haldane gap. Longer chains 
are necessary for the $S = 2$ HAF with $\xi \sim 50$ or for the gapless 
$S = 3/2$ HAF.

In this paper we consider edge states of HAFs with $S = 1$, 3/2 and $2$ 
in systems of up to 500 spins. We use conventional DMRG for chains with 
an even number of spins and another algorithm for chains with an odd 
number of spins. We compute and model the spin densities of edge states 
as well as their excitation energies. The Hamiltonian of the spin-$S$ 
HAF chain with open boundary conditions (OBC) is
\begin{equation}
H_S(N) = J \sum_{r=1}^{N-1} \vec{S}_r \cdot \vec{S}_{r+1}.
\label{eq:ham}
\end{equation}
The spin at site $r$ is $S_r$, the total spin $S_T$ and its $z$ 
component $S^z$ are conserved, and $J = 1$ is a convenient unit of 
energy.  

The terminal spins $r = 1$ and $N$ are coupled to only one spin in 
Eq.~\ref{eq:ham}. Periodic boundary conditions (PBC) also has $J$ 
between sites 1 and $N$. Every spin is then coupled to two neighbors, 
the system has translational symmetry, and the smallest $S_T$ is expected 
in the ground state (GS) for AF exchange. Indeed, the GS of PBC chains 
is a singlet, spin $S_G = 0$, except for odd $N$ and half integer $S$, 
when $S_G = 1/2$. The sectors of integer and half integer $S$ are 
disjoint, and even $N$ is conventionally taken for the thermodynamic 
limit. As noted by Faddeev and Takhtajan, the thermodynamic limit of 
the $S = 1/2$ HAF with odd $N$ is not well understood.~\cite{faddeev81}

HAFs with OBC are fundamentally different because there is no energy 
penalty for parallel spins at sites 1 and $N$. The GS of Eq.~\ref{eq:ham} 
remains a singlet for even $N$, but it becomes a multiplet with 
$S_G = S$ and Zeeman degeneracy $(2S_G + 1)$ for odd $N$. The 
lowest-energy triplet is necessarily an excited state when $N$ is even. 
For integer $S$, the singlet is an excited state when $N$ is odd, while 
for half integer $S > 1/2$, the doublet is an excited state when $N$ is 
odd. Except in the $S = 1/2$ case, $S_G$ depends on the boundary 
conditions for arbitrarily large systems. It follows that HAFs with 
OBC support edge states with $S_G \ge 1$ whose energies become 
degenerate in the thermodynamic limit with those of PBC systems with 
$S_G = 0$ or 1/2. 

We define the energy gaps of edge states in chains of 
$N$ spins as 
\begin{equation}
\Gamma_S(N) = E_0(S,N) - E_0(0,N),
\label{eq:gap}
\end{equation}
where $E_0(S,N)$ is the lowest energy in the sector with total spin $S$. 
Even $N$ leads to $\Gamma_S(N) > 0$. Odd $N$ leads to $\Gamma_S(N) < 0$ for 
integer $S$ and to $\Gamma_S(N) < 0$ relative to $E_0(1/2,N)$ for half 
integer $S$. Since DMRG algorithms conserve $S^z$ rather than $S$, the 
most accurate results are the GS in sectors with increasing $S^z$ and 
$\Gamma_S(N) > 0$. Otherwise, the singlet or doublet is an excited state 
in the $S^z = 0$ sector for integer $S$ or in the $S^z = 1/2$ sector for 
half integer $S$. The size dependence of $\Gamma_S(N)$ is faster than 
$1/N$, which distinguishes gap states from bulk excitations that may 
also have zero gap in the thermodynamic limit. 

We shall characterize edge states using spin densities and call them 
boundary-induced spin density waves (BI-SDWs). BI-SDW is more 
descriptive than edge state and is more accurate than localized state, 
since BI-SDWs are not localized in half integer $S$ chains. By 
convention, we choose the Zeeman level $S^z = S$ when $S \ge 1$ and 
define the spin density at site $r$ as 
\begin{equation}
\rho(r,N) = \langle S^z_r \rangle, \qquad \qquad r=1, 2 \ldots N.
\label{eq:spden}
\end{equation}
The expectation value is with respect to the state of interest. 
Singlet states have $\rho(r,N) = 0$ at all sites. SDWs with $S \ge 1$ 
have equal spin density at $r$ and $N+1-r$ by symmetry in chains, 
$\rho(N) = \rho(1) > 0$ by construction and $\rho(r,N)\propto(-1)^{r-1}$. 
It is advantageous to focus on spin densities rather than energy 
gaps. Spin densities are exclusively associated with $S > 0$ states 
while the $\Gamma_S(N)$ in Eq.~\ref{eq:gap} are small differences 
between extensive energies.

The NL$\sigma$M is a good approximation for $S \ge 1$ HAFs, and 
theoretical discussions have focused as much on field theory as on 
spin chains.~\cite{affleck86a,*affleck86b,schulz86,affleck89} The 
model for integer $S$ chains relates edge states to an effective 
Hamiltonian between spins $s' = S/2$ at the ends,~\cite{machens2013}
\begin{equation}
H_{eff}(N) = (-1)^N J_e\exp(-N/\xi) \vec{s'}_1 
\cdot \vec{s'}_N.
\label{eq:heff}
\end{equation}
The correlation length $\xi$ and exchange $J_e$ are fit to DMRG 
results for $H_S$. An interesting point is that $\xi$ refers to the 
bulk, the singlet GS in the thermodynamic limit, as has been 
confirmed within numerical accuracy in $S = 1$ 
chains.~\cite{white-huse-prb93} The $S = 2$ chain has two gap states 
that afford more stringent tests of Eq.~\ref{eq:heff}. For example, 
the ratio of the two gaps is necessarily 3:1 for $s'_1 = s'_N = 1$. 
Edge states in HAFs with half integer $S \ge 3/2$ have been 
discussed~\cite{tkng94,qin95,machens2013} using $H_{eff}$ 
with effective spins $s' = (S - 1/2)/2$ and effective exchange 
$J'(N)$ that decreases faster than $1/N$ but not exponentially. 

Our principal goal is the quantitative description of edge states in 
HAFs that are sufficiently long to neglect bulk excitations in 
$S = 3/2$ or 2 chains. The paper is organized as follows. 
Section~\ref{sec:dmrg} summarizes conventional DMRG algorithm 
for even $N$ and a different algorithm for odd $N$ that is related 
to $Y$ junctions. Section~\ref{sec:integer} presents BI-SDWs spin 
densities and gaps for $S = 1$ and $S = 2$ chains with finite Haldane 
gaps and finite correlation lengths $\xi$. DMRG returns $\xi = 6.048$ 
for $S = 1$ chains, in agreement with 6.03(1) reported 
previously,~\cite{white-huse-prb93} and $\xi = 49.0$ for $S = 2$ chains. 
DMRG spin densities are fit quantitatively by BI-SDWs that are in 
phase for odd $N$, out of phase for even $N$. The coupling 
$H_{eff}(N)$ between ends is quantitative for $S = 1$ chains 
and is semi quantitative for $S = 2$ chains, in qualitative agreement 
with the VBS picture of localized spins.  Section~\ref{sec:3by2} 
presents the BI-SDWs and gaps of the $S = 3/2$ chain. The BI-SDWs are 
not localized in this case. The singlet-triplet gap $\Gamma_1(N)$ for 
even $N$ decreases faster than $1/N$, as anticipated by Ng.~\cite{tkng94} 
The gap $\Gamma_{3/2}(N)$ for odd $N$ requires a modified 
$H_{eff}(N)$ with a delocalized spin in central part in 
addition to spins at the ends. The delocalized spin rationalizes 
$|\Gamma_{3/2}(N)| > \Gamma_1(N)$ and a weaker size dependence. The 
Discussion summarizes the limited nature of connections to the 
NL$\sigma$M or to VBS. 

\section{\label{sec:dmrg}DMRG algorithms}
By now, DMRG is a mature numerical method for 1D 
systems.~\cite{schollwock2005,karen2006} It gives excellent GS 
properties and has been widely applied to spin chains. 
Conventional DMRG starts with a superblock that consists of four 
sites: one site in the left block, one in the right block and two 
new sites, the central sites. The left and right blocks increase 
by one site as two new sites are added at every step. The procedure 
generates a chain with OBC and an even number of sites $N$. The 
vast majority of DMRG calculations been performed on chains with 
even $N$. White has discussed~\cite{white-odd-n} an algorithm with 
one rather than two central sites that speeds up the computational 
time by a factor of two to four. The method was tested on an $S = 1$ 
HAF of 100 spins.  

We use conventional DMRG for spin chains with even $N$ and adapt an 
algorithm for odd $N$ that was developed for $Y$ junctions.~\cite{mk2016} 
$Y$ junctions of $N = 3n + 1$ spins have three arms of $n$ spins plus 
a central site for which we recently presented an efficient DMRG 
algorithm. Fig. 2 of Ref.~\onlinecite{mk2016} shows the growth of the 
infinite DMRG algorithm. A chain of $N = 2n + 1$ spins can be viewed as 
two arms of $n$ spins plus a central site. The algorithm takes the 
system as an arm plus the central site and the environment as the other 
arm. Since the system of $n + 1$ spins at step $n$ becomes the 
environment at the next step, the chain grows by two spins at each 
step. The procedure described for $Y$ junctions~\cite{mk2016} holds 
with one fewer arm for chains of $N = 2n + 1$ spins.

The accuracy of the algorithm for odd $N$ is comparable to conventional 
DMRG, as has already been shown for $Y$ functions.~\cite{mk2016} In 
either algorithm, new sites are coupled to the most recently added 
sites and the superblock Hamiltonian contains only new and once 
renormalized operators. Table~\ref{tab:gs} has representative DMRG 
results for the ground states of $S = 2$ and 3/2 chains with $N \ge 100$ 
spins. The index $m$ is the number of states kept per block. The 
truncation error is $P(m) = 1 - \sum_j^m \omega_j$ where the sum is 
over the eigenvalues $\omega_j$ of the density matrix. Several sweeps 
of finite DMRG calculations are required for $S = 3/2$ or 2, with $N$ 
calculations per sweep, and finite DMRG is necessary for accurate 
spin densities. Increasing $m$ rapidly increases the required computer 
resources for long chains and involves trade offs. We have checked our 
results against previous studies in Table~\ref{tab:gs} as well as 
against $S = 1$ chains and find comparably small or smaller $P(m)$ that 
amount to evolutionary improvements for even $N$. The algorithm for odd 
$N$ returns equally small $P(m)$. 
\begin{table}
\caption{\label{tab:gs}Representative previous and present DMRG
calculations for HAF chains with spin $S = 2$ or 3/2 and
$N \ge 100$ sites. The truncation error is $P(m) = a \times 10^{-b}$
when $m$ states are kept per block.}
\begin{ruledtabular}
\begin{tabular}{ c  c  c  c } 
N, [Ref.]  & $S$  & $m$  &  $P(m)$   \\ \hline
270 [\onlinecite{schollwock96}] & 2 & 210 & 1E-7 \\
150 [\onlinecite{qin97}] & 2 & 250 & 1E-6 \\
100 [\onlinecite{qin95}] & 3/2 & 120 & 3E-6 \\
192 [\onlinecite{fath2006}] & 3/2 & 500--800 & 1E-7 -- 1E-8  \\
400  & 2  & 600  & 1.1E-7  \\
399  & 2  & 500  & 8.5E-9  \\
400  & 3/2 & 460 & 8.4E-8  \\
399 & 3/2 & 460 & 1.4E-8 \\
\end{tabular}
\end{ruledtabular}
\end{table}

In the following we have set $m$ according to Table~\ref{tab:gs} and 
performed 5-10 sweeps of finite DMRG for $S = 2$ and 3/2 chains. 
We estimate that GS energies per site are accurate to $10^{-8}$ for 
$S = 1$ chains, to $10^{-6}$ for $S = 3/2$ and to $10^{-5}$ for 
$S = 2$. The energy gaps $\Gamma_S(N)$ between the GS in sectors with 
different total spin are accurate to $10^{-5}$ for $S = 1$ and to 
$10^{-4}$ for $S =  3/2$ or 2. The spin densities are estimated to be 
accurate to better than $10^{-4}$ based, for example, on DMRG calculations with 
different algorithms for $N$ and $N - 1$. Accurate $\rho(r,N)$ are 
readily obtained in large systems whose $\Gamma_S(N)$ are not accessible.

\section{\label{sec:integer}Integer spin, $\mathbf{S = 1}$ and 2}
\begin{figure}
\includegraphics[width=\columnwidth]{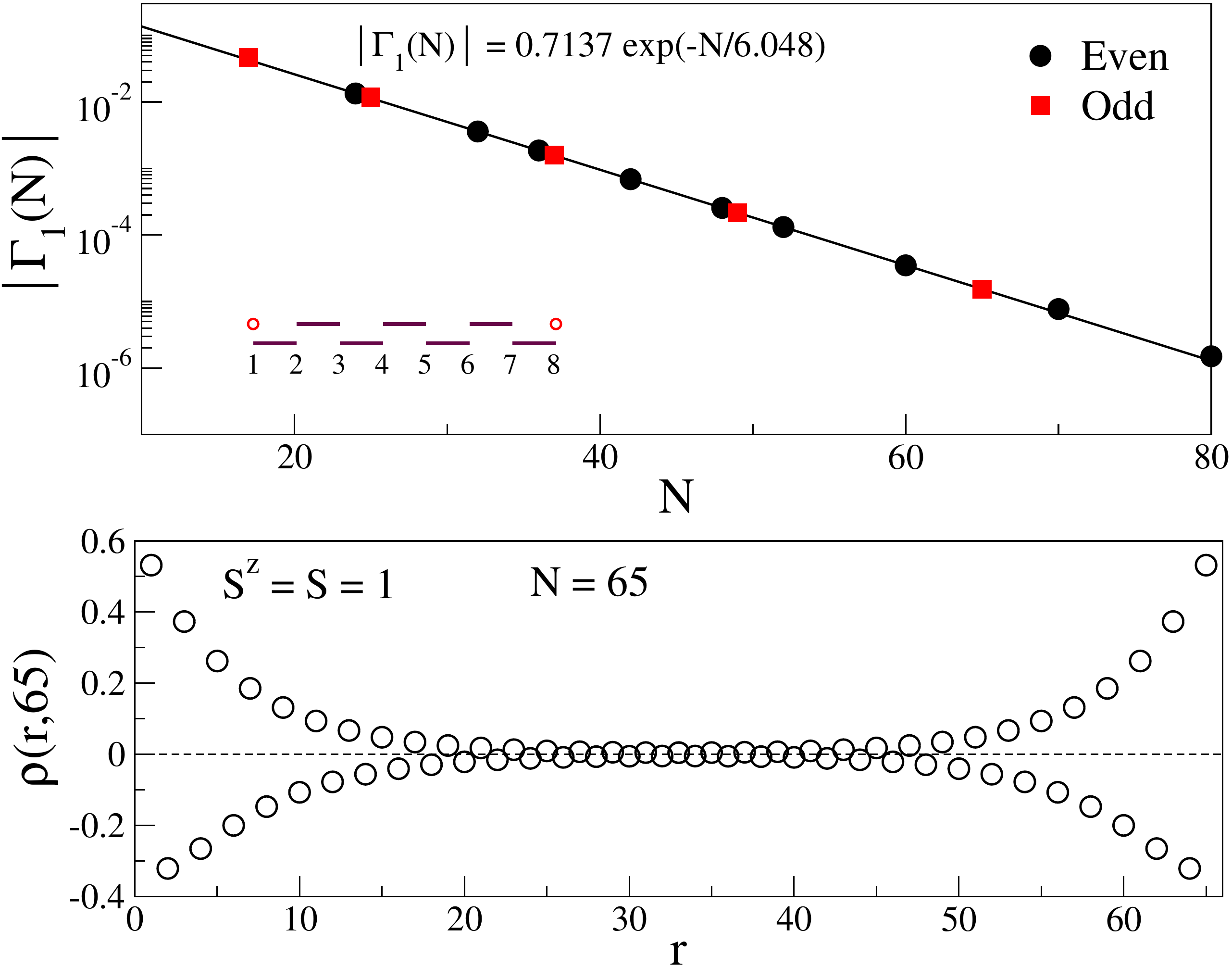}
\caption{\label{fig:gap-s1}Upper panel: Singlet-triplet gap
$|\Gamma_1(N)|$ of $S = 1$ chains with $N$ spins, even or odd, 
in Eq.~\ref{eq:ham};
inset: VBS valence bond diagram for $N = 8$. Lower panel:
Spin densities $\rho(r,65)$ of the $S = 1$ chain with 65 spins.}
\end{figure}

We start with the extensively studied $S = 1$ HAF with OBC and even $N$. 
The large Haldane gap~\cite{white-huse-prb93,nakano2009} 
$\Delta(1) = 0.4105$ reduces the computational effort. The singlet-triplet 
gap $\Gamma_1(N)$ in Eq.~\ref{eq:gap} decreases rapidly with system size. 
The GS alternates between $S_G = 0$ and 1 for even and odd $N$, 
respectively. We evaluate $\Gamma_1(N)$ for even $N$ as the difference of 
the total energy in the sectors $S^z = 0$ and 1. In addition, we also 
obtain $\Gamma_1(N) < 0$ for odd $N$ using the first excited state in 
the $S^z = 0$ sector. The excited state is accurate to $10^{-6}$ for 
$m > 300$. As shown in Fig.~\ref{fig:gap-s1}, upper panel, with different
symbols for even and odd $N$, $|\Gamma_1(N)|$ decreases as 
$J_e \exp(-N/\xi)$ with $\xi = 6.048$. The effective exchange between the ends 
is $J_e = 0.7137$ in Eq.~\ref{eq:heff} with spins $s' = 1/2$. The 
effective Hamiltonian is quantitative for $S = 1$ chains.  The gap 
at $N = 80$ is $1.5 \times 10^{-6}$, which still exceeds the estimated numerical 
accuracy. The inset shows the relevant VBS valence bond 
diagram.~\cite{affleck88} Each line is a singlet pair, 
$(\alpha \beta - \beta \alpha)/\sqrt{2}$, between $S =1/2$ spins, 
two per $S = 1$ site, and the circles are unpaired spins at the ends. 

Comparable DMRG accuracy for $S = 1$ chains with even $N$ has been 
discussed previously. S$\o$rensen and Affleck found $\xi = 6.07$ for 
$\Gamma_1(N)$ and 6.028(3) for spin densities.~\cite{sorensen94} 
White and Huse obtained~\cite{white-huse-prb93} the GS energy per site 
very accurately and reported $\xi = 6.03(1)$ for the spin densities 
of a 60-site chain with 
an auxiliary spin-1/2 at one end (site $N + 1$ in Eq.~\ref{eq:ham}). 
Schollw\"ock et al.,~\cite{schollwock96} discussed the same procedure 
for $S = 2$ chains with even $N$ and an auxiliary spin-1 at one end. 
Auxiliary spins at both ends with adjustable exchange to sites 1 and 
$N$ can be used to study bulk excitations.~\cite{white-huse-prb93} 
In this paper, we shall not resort to auxiliary spins. We always 
consider BI-SDWs at both ends of chains.

The spin densities $\rho(r,65)$ in Fig.~\ref{fig:gap-s1}, lower panel, 
are for the GS of the 65-spin chain. We take $S^z = 1$ and obtain 
positive $\rho(r)$ at odd numbered sites and negative $\rho(r)$ at 
even numbered sites, respectively. All chains with $S > 0$ have 
$\rho(r,N) \propto (-1)^{r-1}$, which is why call them as BI-SDWs. 
Table~\ref{tab:spden} lists the spin densities of the first 10 sites 
in chains of 66/65 spins and 48/47 spins. The 66/65 spin densities 
clearly refer to the same triplet and speak to the numerical accuracy 
since different algorithms are used. The spin density at site 1 is 
slightly greater than 1/2, and so is the total spin density to 
odd-numbered sites. The total spin density to an even-numbered site 
approaches 1/2 from below and exceeds 0.45 at $r = 10$. The apparent 
exponential decrease of $|\rho(r,N)|$ does not hold for the first 
few sites since, for example, $|\rho(2)| < \rho(3)$. The triplets 
are identical near the ends but of course differ at the middle of 
the chain, where $\rho(33,65) = 4.82 \times 10^{-3}$ becomes 
$\rho(33,66) = \rho(34,66) = 3.7 \times 10^{-4}$. The 48/47 data 
illustrate the weak size dependence of spin densities at the ends. 
Well-defined edge states must become size independent. The first 
10 sites of $N = 65$ or 66 chains are close to the thermodynamic 
limit of BI-SDWs.

\begin{figure}
\includegraphics[width=\columnwidth]{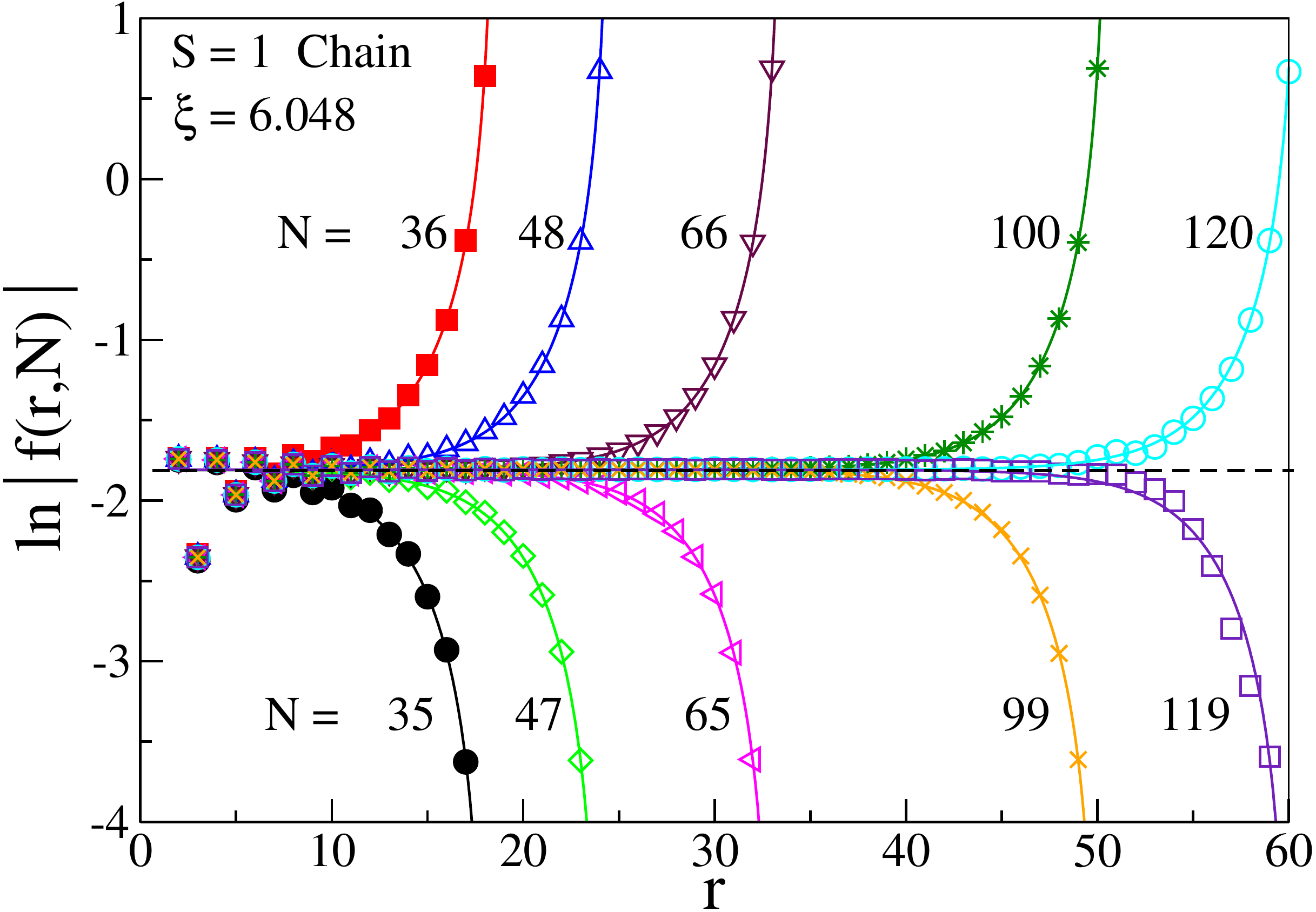}
\caption{\label{fig:lnfr-s1} Symbols are DMRG spin densities in 
$f(r,N)$, Eq.~\ref{eq:fdef}, to the middle of $S = 1$ chains of $N$ 
spins. The lines are Eq.~\ref{eq:fr-post} with correlation length 
$\xi = 6.048$. The horizontal dashed line is the thermodynamic limit.}
\end{figure}

\begin{table}
\caption{\label{tab:spden}DMRG results for spin densities at the
first ten sites of $S = 1$ chains of $N$ spins.}
\begin{ruledtabular}
\begin{tabular}{ c  c  c  c  c}
Site  & N = 66 & N = 65 & N = 48 & N = 47  \\ \hline
1 & 0.53204 & 0.53204 & 0.53198 & 0.53211  \\
2 & -0.32090 & -0.32091 & -0.32081 & -0.32102 \\
3 &  0.37324 & 0.37326 & 0.37311 & 0.37342  \\
4 &  -0.26515 & -0.26517 & -0.26495 & -0.26541 \\
5 & 0.26242 & 0.26245 & 0.26216 & 0.26275 \\
6 & -0.19992 & -0.19996 & -0.19958 & -0.20037 \\
7 & 0.18544 & 0.18549 & 0.18502 & 0.18599 \\
8 & -0.14688 & -0.14694 & -0.14635  & -0.14757 \\
9 & 0.13166 & 0.13173 & 0.13101  & 0.13249  \\
10 & -0.10675 & -0.10685 & -0.10597 & -0.10777 \\
\end{tabular}
\end{ruledtabular}
\end{table}
To minimize the even-odd variations of spin densities and to divide 
out an overall scale factor, we consider the function
\begin{equation}
f(r,N) = \frac{\rho(r-1)-\rho(r+1)}{\rho(r-1)+\rho(r+1)} 
     \approx -\frac{\partial}{\partial r} \ln \rho(r,N).
\label{eq:fdef}
\end{equation}
$f(r,N)$ is odd with respect to the chain's midpoint while $\rho(r,N)$ 
is even. Figure~\ref{fig:lnfr-s1} shows $\ln |f(r,N)|$ for $S = 1$ 
chains up to the middle, $r \le (N + 1)/2$. The DMRG points near the 
edge become size independent. Except for the first few ($\sim 10$) sites, 
$f(r,N)$ is constant up to about $N/2 - 2\xi$. The difference between even 
and odd $N$ is clearly seen in the middle region, and $f(r,N)$ for 
even, odd pairs are a convenient way to present spin densities directly 
without making any assumptions about the appropriate model or 
interpretation. It follows that the thermodynamic limit is 
$\ln |f(r)| = -1.801$. The lines are fits as discussed below using 
the correlation length $\xi = 6.048$ from the gap $\Gamma_1(N)$, in 
accord with the NL$\sigma$M's expectation of equal $\xi$ for gaps 
and spin densities.

\begin{figure}
\includegraphics[width=\columnwidth]{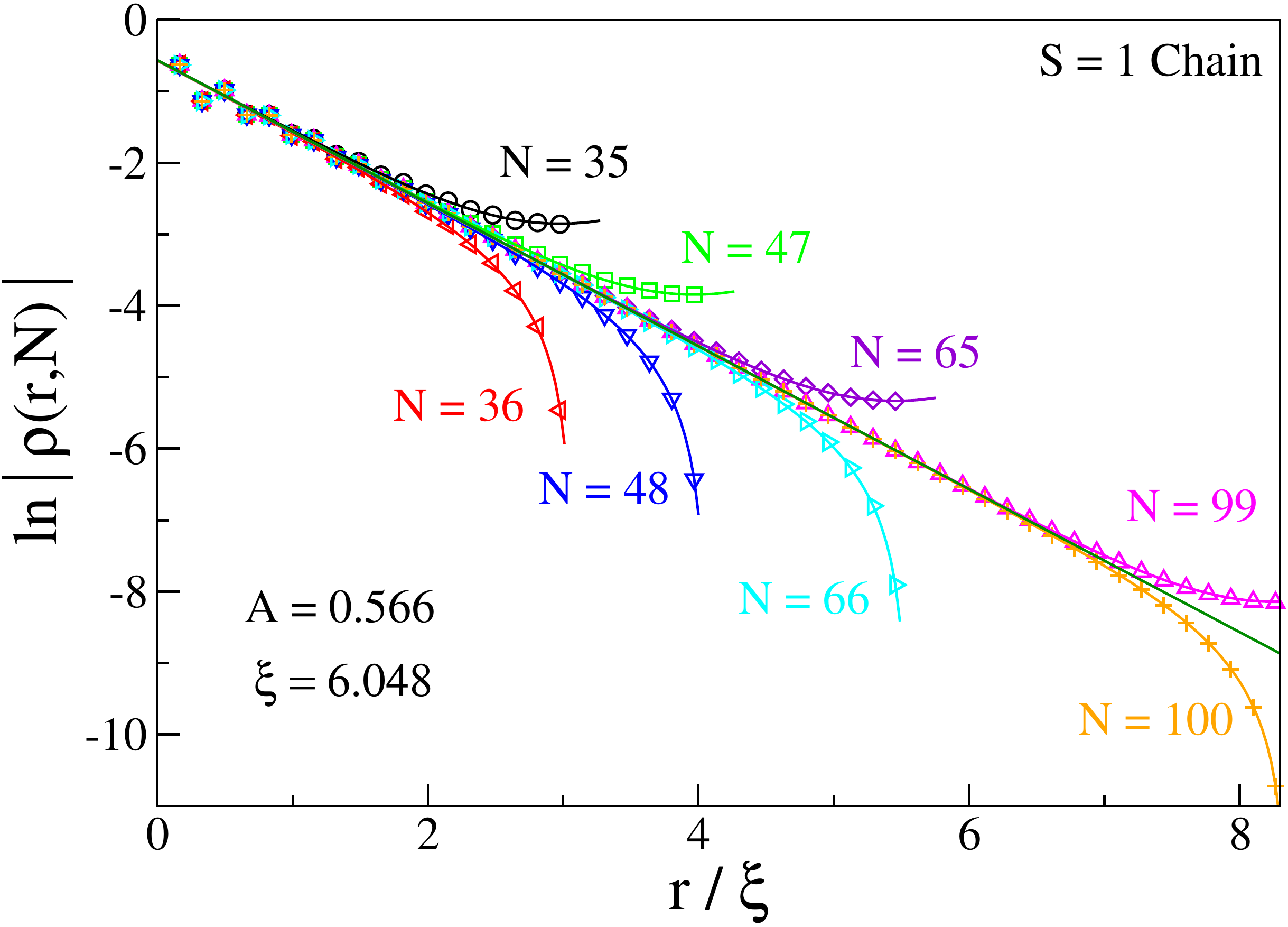}
\caption{\label{fig:amp-s1} Symbols are DMRG results for $|\rho(r,N)|$ 
to the middle of $S = 1$ chains. Lines are Eq.~\ref{eq:spden-post} with 
$\xi = 6.048$ and $A = 0.566$. Even and odd $N$ deviate from 
$A \exp(-r/\xi)$ near the middle of chains.}
\end{figure}

The magnitudes of the spin densities are shown in Fig.~\ref{fig:amp-s1} 
as a function of $r/\xi$ up to the middle of the chains. They decrease 
as $A \exp(-r/\xi)$ and deviate upward in the middle for odd $N$, 
downwards for even $N$. The amplitude $A$ is independent of system size when 
$N/\xi > 5$.

To model the spin densities of integer $S$ chains, we introduce SDWs 
at the left and right ends,
\begin{eqnarray}
\rho(r,N) &=& A (-1)^{r-1} \left[ \exp(-r/\xi) \right. \nonumber \\
& & \left. - (-1)^N \exp \left(-(N+1-r)/\xi\right) \right].
\label{eq:sdw}
\end{eqnarray}
The SDWs are in phase for odd $N$ when all odd-numbered sites have 
$\rho > 0$; they are out of phase for even $N$ with equal $\rho$ at 
sites $N/2$ and $N/2 + 1$. Except for Ref.~\onlinecite{sorensen94}, 
the spin densities have been assumed to decrease exponentially, 
thereby ignoring contributions from the other end. While that is the 
case in the thermodynamic limit, $N > 10\xi$ is minimally required to 
neglect contributions from the other BI-SDW in the middle. 
Since the system size in DMRG 
calculations rarely exceeds $10 \xi$, it is advantageous to consider 
both ends. We have
\begin{widetext}
\begin{equation}
\rho(r,N) = 2 A (-1)^{r-1} \exp \left(-(N+1)/2\xi\right) 
\begin{cases}
\cosh \left((N+1-2r)/2\xi \right), & \quad ($odd $ N) \\
\sinh \left((N+1-2r)/2\xi \right), & \quad ($even $ N) \\
\end{cases}
\label{eq:spden-post}
\end{equation}
The postulated BI-SDWs lead to
\begin{equation}
f(r,N) = \tanh(1/\xi)
\begin{cases}
\tanh \left( (N+1-2r)/2\xi \right), \quad ($odd $ N) \\
\coth \left( (N+1-2r)/2\xi \right), \quad ($odd $ N) \\
\end{cases}
\label{eq:fr-post}
\end{equation}
\end{widetext}
The relative phase of the SDWs matters within $\pm 2\xi$ of the 
middle. The range of $r$ is the same for $N$ and $N - 1$ when 
$N$ is even.  

The lines in Fig.~\ref{fig:lnfr-s1} are $\ln |f(r,N)|$ for 
$\xi = 6.048$ and continuous $r$ in Eq.~\ref{eq:fr-post}. The 
thermodynamic limit is $f(r) = \tanh(1/\xi)$, the dashed line in 
Fig.~\ref{fig:lnfr-s1}, and it reduces to $1/\xi$ for a continuous 
chain. The correlation length $\xi$ is accurately obtained using both 
even and odd chains. The $N = 119/120$ spin densities indicate a gap 
of $\Gamma_1(120) = 1.72 \times 10^{-9}$ that is far below the 
accuracy of the energy difference. DMRG spin densities are also 
limited, however, to less than 149/150; there the $\rho(r,N)$ show 
considerable scatter where $f(r,N)$ has even-odd variations. The SDW 
amplitude $A = 0.656$ in Fig.~\ref{fig:amp-s1} accounts quantitatively 
for spin densities aside from the first few. The parameters $\xi$ 
and $A$ suffice for all fits in 
Figs.~\ref{fig:lnfr-s1}~and~\ref{fig:amp-s1}. 

The $S = 2$ chain has a smaller Haldane gap~\cite{nakano2009} of 
$\Delta(2) = 0.088$. Numerical analysis is more difficult since (i) 
there are more degrees of freedom per site; (ii) $N > 5\xi$ requires 
longer chains; and (iii) gaps $\Gamma_S(N) < \Delta(2)$ also require 
longer chains to distinguish between edge and bulk excitations. Results 
are fewer and less accurate. The nature of BI-SDWs in $S = 2$ or 3/2 
chains was the motivation for DMRG calculations on even and odd chains 
of hundreds of spins. According to the NL$\sigma$M, edge states for 
$S = 2$ are associated with spin $s' = S/2 = 1$ in $H_{eff}$, 
Eq.~\ref{eq:heff}. Even chains have a singlet GS and gaps to two edge 
states, $\Gamma_1(N)$ to the triplet ($S = 1$) and $\Gamma_2(N)$ to the 
quintet ($S = 2$). The correlation length $\xi$ is the same for both 
and $\Gamma_2(N) = 3 \Gamma_1(N)$. The VBS valence bond diagram 
corresponds to two $S = 1$ diagrams in Fig.~\ref{fig:gap-s1}(a): 
There are four lines per interior $S = 2$ site and two lines, two 
unpaired spin at the ends. The BI-SDW analysis of $S = 1$ chains is 
equally applicable to integer $S$ chains. Increasing $S$ leads to 
longer $\xi$ and to gaps $\Gamma_S(N)$ whose relative magnitudes are 
fixed in advance by Eq.~\ref{eq:heff}. 

A chain with $J = 1$ and 200 spins $S = 2$ or 400 spins $S = 3/2$ has 
a GS energy of roughly $-10^3$. The corresponding $\Gamma(N)$ in 
Table~\ref{tab:gaps} are less than $10^{-3}$ and their estimated 
accuracy is $\pm 1\times 10^{-4}$. Our $S = 2$ and 3/2 gaps are 
consequently limited to $N \sim 200$ and 450, respectively. They 
are differences between total energies. Spin densities, by contrast, 
are exclusively related to the GS in a sector with $S > 0$. The 
representative gaps in Table~\ref{tab:gaps} cover more than a decade.  
We studied the $m$ dependence of gaps in $S = 2$ and 3/2 chains, 
summarized in Table~\ref{tab:gs}, in order to find the largest 
accessible systems. The gap $\Delta(2)$ of the infinite $S = 2$ 
chain is slightly larger than $\Gamma_2(32)$. The competition 
between edge and bulk excitations in short HAFs with $S \ge 1$ is 
discussed elsewhere.~\cite{qin95,fath2006,machens2013,lou2002}

\begin{table}
\caption{\label{tab:gaps}Edge-state energy gaps $\Gamma(N)$,
Eq.~\ref{eq:gap}, of HAFs with $N$ spins $S$ and $J = 1$ in
Eq.~\ref{eq:ham}.}
\begin{ruledtabular}
\begin{tabular}{ c  c  c  c}
N & $\Gamma_1(N)$, $S = 2$ & $\Gamma_2(N)$, $S = 2$ & $\Gamma_1(N)$, $S = 3/2$
\\ \hline
64 & 0.01583 & 0.05578 & 0.01276 \\
100 & 0.00721 & 0.02452 & 0.00725 \\
150 & 0.00255 & 0.00835 & 0.00388 \\
200 & 0.00081 & 0.00285 & 0.00268 \\
300 & -- & -- & 0.00194 \\
400 & -- & -- & 0.00090 \\
450 & -- & -- & 0.00065 \\
\end{tabular}
\end{ruledtabular}
\end{table}

\begin{figure}
\includegraphics[width=\columnwidth]{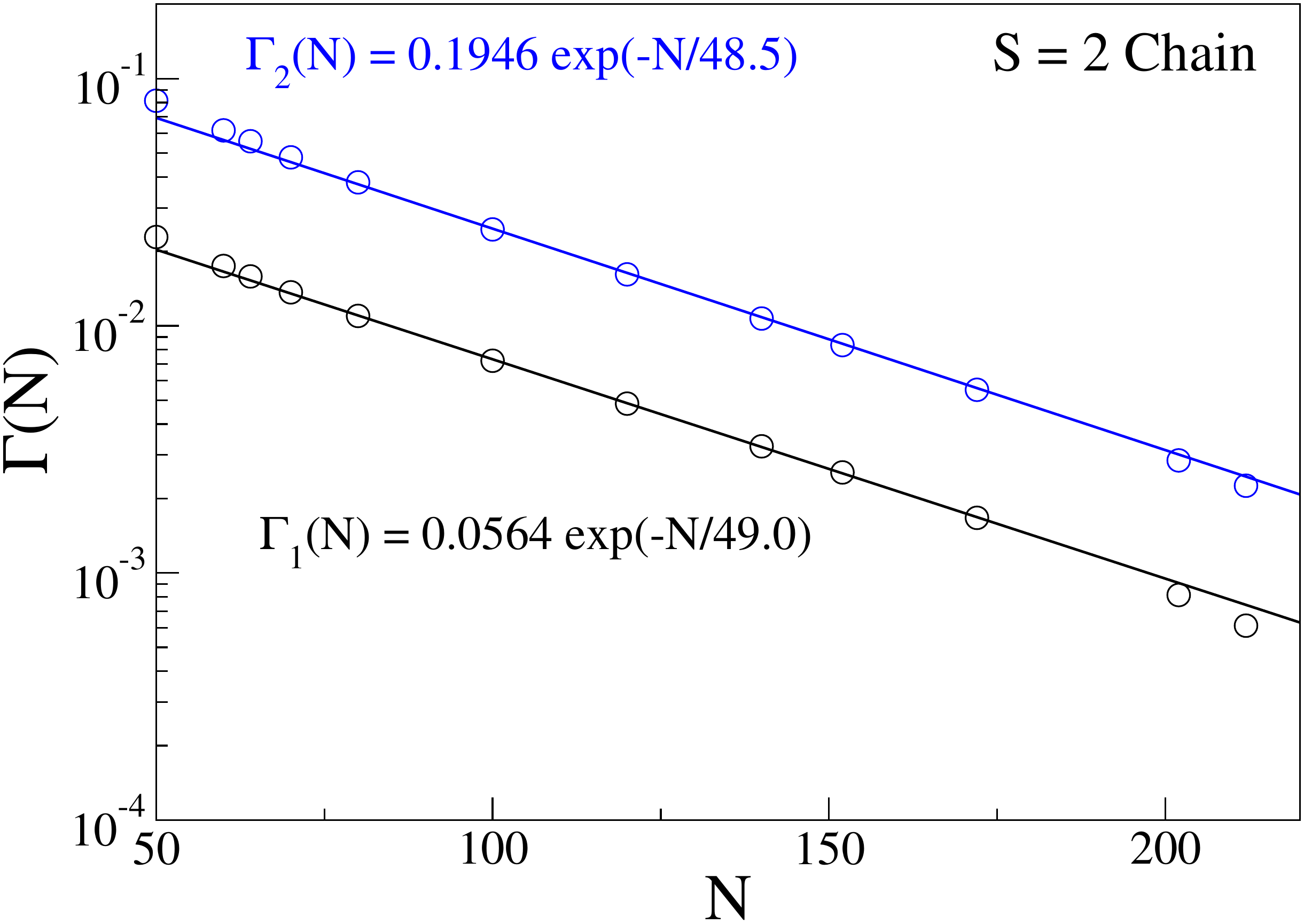}
\caption{\label{fig:gap-s2}Edge-state gaps $\Gamma_1(N)$ and
$\Gamma_2(N)$ of $S = 2$ chains of $N$ spins in Eq.~\ref{eq:ham}.}
\end{figure}

Figure~\ref{fig:gap-s2} shows $\Gamma_1(N)$ and $\Gamma_2(N)$ for 
$S = 2$ and even $N$. The gaps are exponential in $N/\xi$, as expected 
for integer $S$. The correlation length, $\xi \sim 49$, is the same 
within our numerical accuracy. The ratio is $\Gamma_2/\Gamma_1 = 3.45$ 
based on the fitted lines and it varies between 3.27 and 3.56 for 
individual points. Although $\Gamma_2/\Gamma_1 = 3.45$ is approximate, 
the ratio is larger than the NL$\sigma$M value of 3 based on 
Eq.~\ref{eq:heff}. We return to gaps after presenting results for spin 
densities.

\begin{figure}
\includegraphics[width=\columnwidth]{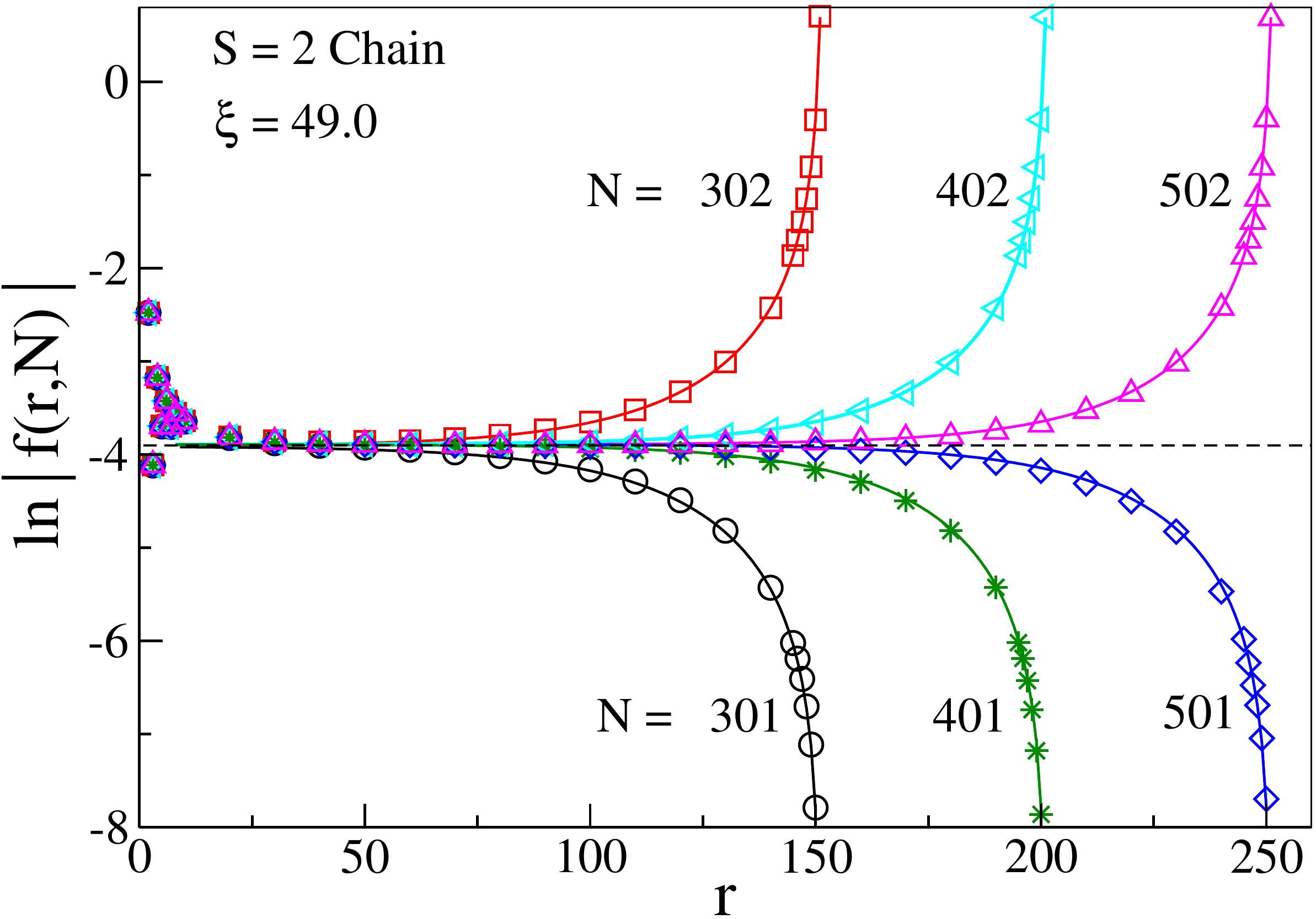}
\caption{\label{fig:lnfr-s2}Open symbols are DMRG spin densities
in the $S^z = 2$ sector for $f(r,N)$, Eq.~\ref{eq:fdef}, to the
middle of $S = 2$ chains of $N$ spins. Solid lines are Eq.~\ref{eq:fr-post}
with correlation length $\xi = 49.0$. The horizontal dashed
line in the thermodynamic limit.}
\end{figure}

\begin{figure}
\includegraphics[width=\columnwidth]{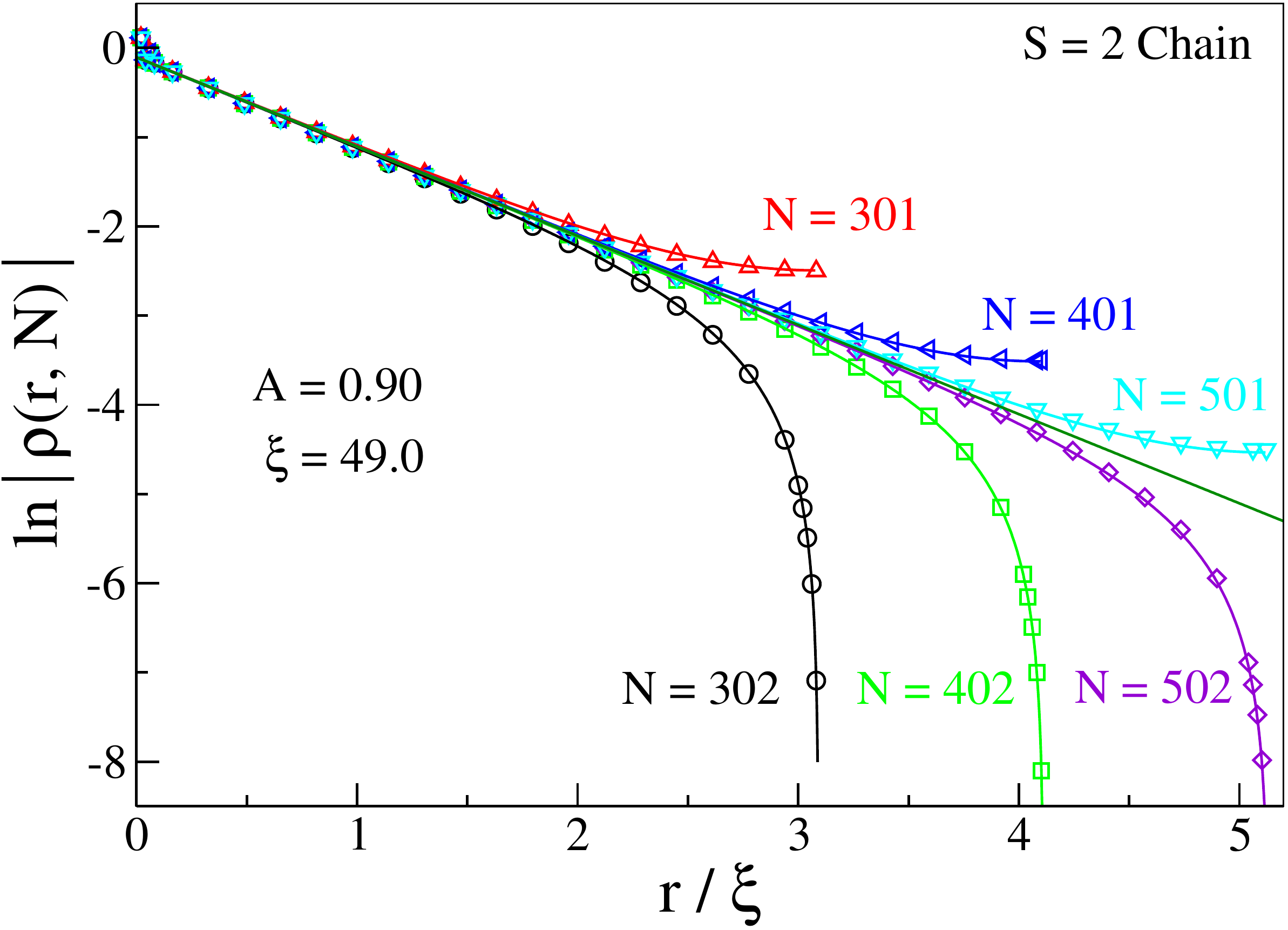}
\caption{\label{fig:amp-s2}Open symbols are DMRG results for
$|\rho(r,N)|$ in the $S^z = 2$ sector to the middle of $S = 2$
chains. Lines are Eq.~\ref{eq:fr-post} with $\xi = 49.0$
and $A = 0.90$. Even and odd $N$ deviate from $A \exp(-r/\xi)$
in the middle.}
\end{figure}

$S = 2$ chains with odd $N$ have a quintet GS and excitations to the 
triplet and singlet. We again use $f(r,N)$ and the BI-SDWs analysis. 
Figure~\ref{fig:lnfr-s2} shows $\ln |f(r,N)|$ in the $S^z = 2$ sector 
up to the middle of the chains. For the sake of clarity, not all 
points are shown. Even-odd effects now extend to about the first 25 
sites and become size independent in long chains. The thermodynamic 
limit is $f(r) = \tanh(1/\xi)$ with $\xi = 49.0$. The magnitudes of 
spin-densities in Fig.~\ref{fig:amp-s2} are fit as a function of 
$r/\xi$ with the same $\xi$ and $A = 0.90$ in Eq.~\ref{eq:spden-post}. 
Two parameters are nearly quantitative aside from sites $r < 25$. 
The triplet is an excited state for either even or odd $N$. It is the 
lowest state in the $S^z = 1$ sector for even $N$ and the first excited 
state in that sector for odd $N$. DMRG calculations for even $N$ converge 
slowly for reasons we do not understand in detail. The triplet spin 
densities return the same $f(r,N)$ as the quintets in Fig.~\ref{fig:lnfr-s2}.

The correlation length $\xi = 49.0$ based on spin densities is more 
accurate than $\xi$ from energy gaps. The $\xi = 49.0$ fits account 
for $\rho(r,N)$ of even and odd chains that extend to 500 spins, 
whereas numerical accuracy limits $\Gamma_S(N)$ to $N \sim 200$. 
Schollw\"ock et al.,~\cite{schollwock96} argued that the thermodynamic 
limit requires $N > 5\xi$ and obtained (Fig. 6 of [\onlinecite{schollwock96}])
$\xi = 49(1)$ for $N = 270$ with an auxiliary spin-1 at the other 
end using the local correlation length $\xi(r) = 2/[ln(\rho(r-1)/\rho(r+1))]$. 
Qin et al.,~\cite{qin95} estimated that $\xi \sim 33$ for S = 2 chains 
up to $N = 100$ and remarked that the accuracy was much worse than 
for $S = 1$ chains. Indeed, spin densities for $N = 127/128$ return 
$\xi \sim 36$. As seen in Figs.~\ref{fig:amp-s1}~and~\ref{fig:amp-s2} 
for $S = 1$ and 2, respectively, the thermodynamic limit requires 
$N > 5\xi$ even when the contribution of the BI-SDW at the other end 
is included. The present results for $S = 2$ chains offer more stringent 
comparisons of the NL$\sigma$M. The model is semi quantitative: 
The ratio $\Gamma_2/\Gamma_1 = 3.45$ is greater than 3. We note that 
BI-SDWs with exponentially decreasing $\rho(r,N)$ would assumed on 
general grounds and follows directly from $f(r, N)$, 
but not the same $\xi$ for gaps and spin densities.  

\section{\label{sec:3by2}Half integer spin, $\mathbf{S = 3/2}$}
HAF chains with half integer $S \ge 3/2$ are gapless and their edge 
states are fundamentally different. Even chains have a singlet GS 
and BI-SDWs with integer $S$; odd chains have $S_G = S$ and BI-SDWs 
with half integer $S > 1/2$. The even $S = 3/2$ chain has a 
singlet-triplet gap $\Gamma_1(N)$ that decreases faster than $1/N$ 
and has been studied by Qin et al.,~\cite{qin95} and in greater 
detail by F\'ath et al.~\cite{fath2006}  The NL$\sigma$M gap goes 
as~\cite{fath2006} 
\begin{equation}
N \Gamma_1(N) = \frac{a}{\ln B N} + O\left( \frac{\ln \ln N}{(\ln N)^2} \right)
\label{eq:fath}
\end{equation}
Fath et al.~\cite{fath2006} used DMRG to compute $\Gamma_1(N)$ for 
$S = 3/2$ chains from $N = 12$ to $192$ in steps of 12 spins. The 
first term of Eq.~\ref{eq:fath} leads to parameters $a(N)$ and $B(N)$ 
whose size dependence was obtained from successive gaps 
$\Gamma_1(N+12)$ and $\Gamma_1(N)$. Extrapolation in $1/N$ gave the 
thermodynamic values of $a = 1.58$ and $B = 0.11$ with $\pm 15\%$ 
uncertainties. In the present study, we are characterizing BI-SDWs 
in spin chains and take the first term with constant $a$, $B$ as a 
two-parameter approximation.

\begin{figure}
\includegraphics[width=\columnwidth]{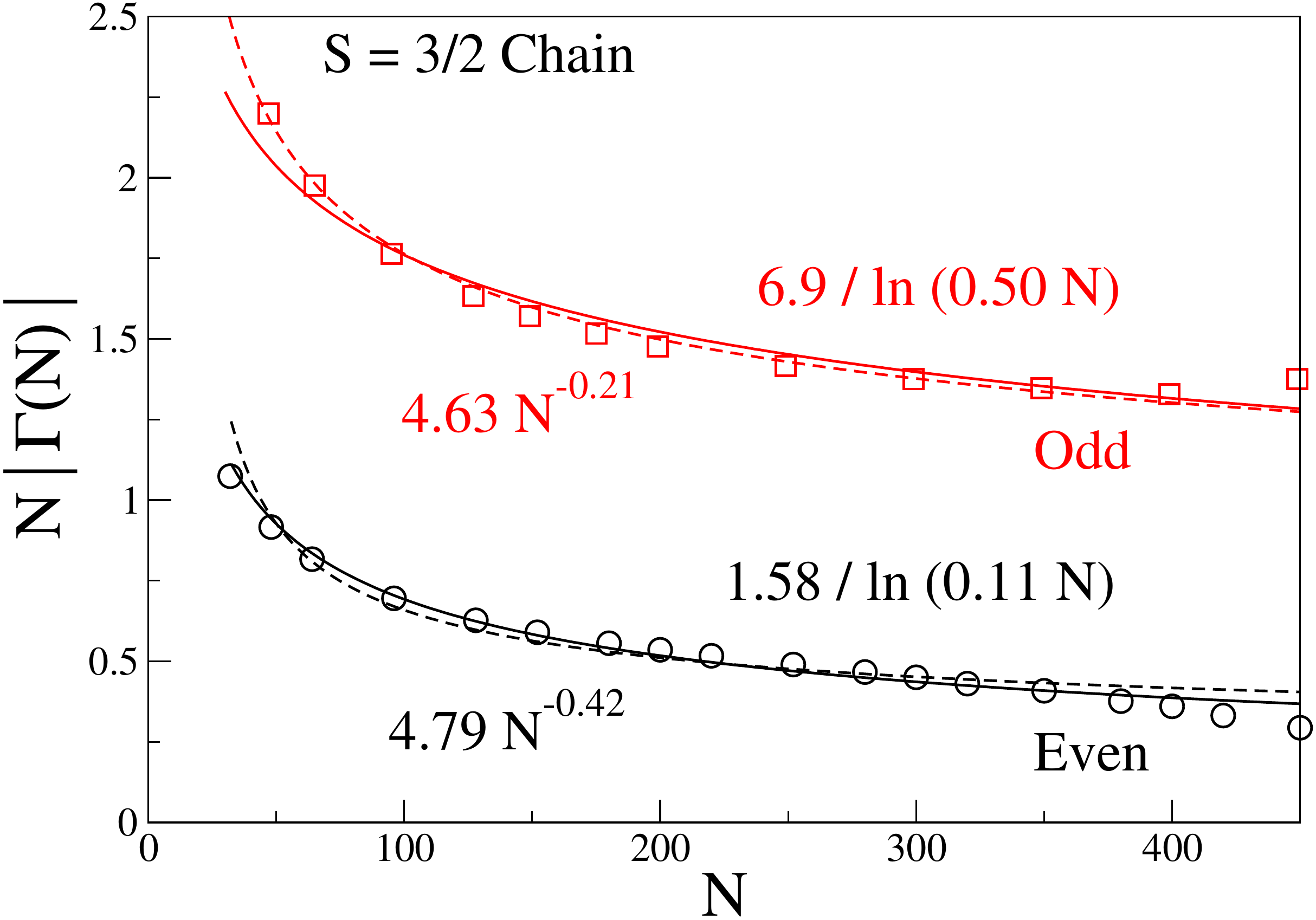}
\caption{\label{fig:gap-s3by2} Open symbols are DMRG results for 
the gaps $\Gamma_1(N)$ and $|\Gamma_{3/2}(N)|$ for $S = 3/2$ 
chains with $N$ spins in Eq.~\ref{eq:ham}. The solid and dashed 
lines are power law and logarithmic fits, respectively, with two 
parameters. For even $N$, the NL$\sigma$M parameters in Eq.~\ref{eq:fath} 
are $a = 1.58$, $B = 0.11$.}
\end{figure}

Figure~\ref{fig:gap-s3by2} shows the calculated gaps of $S = 3/2$ 
HAFs as $N|\Gamma(N)|$. The gaps decrease faster than $1/N$ as expected 
for edge states. The NL$\sigma$M size dependence for even $N$ is 
Eq.~\ref{eq:heff} with $J_e(N) = \Gamma_1 = a/(N \ln B N)$. The dashed 
line has $a = 1.58$ and $B = 0.11$ as inferred by F\'ath 
et al.~\cite{fath2006} The solid line for even $N$ is a power law with 
two parameters, $\Gamma_1 = J_e(N) = 4.79 N^{-1.42}$. Either fit is 
adequate over this range of system sizes, and neither accounts the 
decrease at $N > 400$. The shortest chains in which edge and bulk 
excitations are decoupled are probably in the range $N = 30$ to 60, 
and the desired $\Gamma_1(N)$ fits are for long chains. The gaps 
$|\Gamma_{3/2}(N)|$ for odd $N$ are several times larger and 
their size dependence is weaker. They can be approximation by a 
different logarithm or power law. The gaps $\Gamma_{3/2}(N)$ 
and $\Gamma_1(N)$ of the $S = 3/2$ chain are in marked contrast 
to equal $|\Gamma_1(N)|$ in Fig.~\ref{fig:gap-s1} for $S = 1$ 
chains with even and odd $N$.

\begin{figure}
\includegraphics[width=\columnwidth]{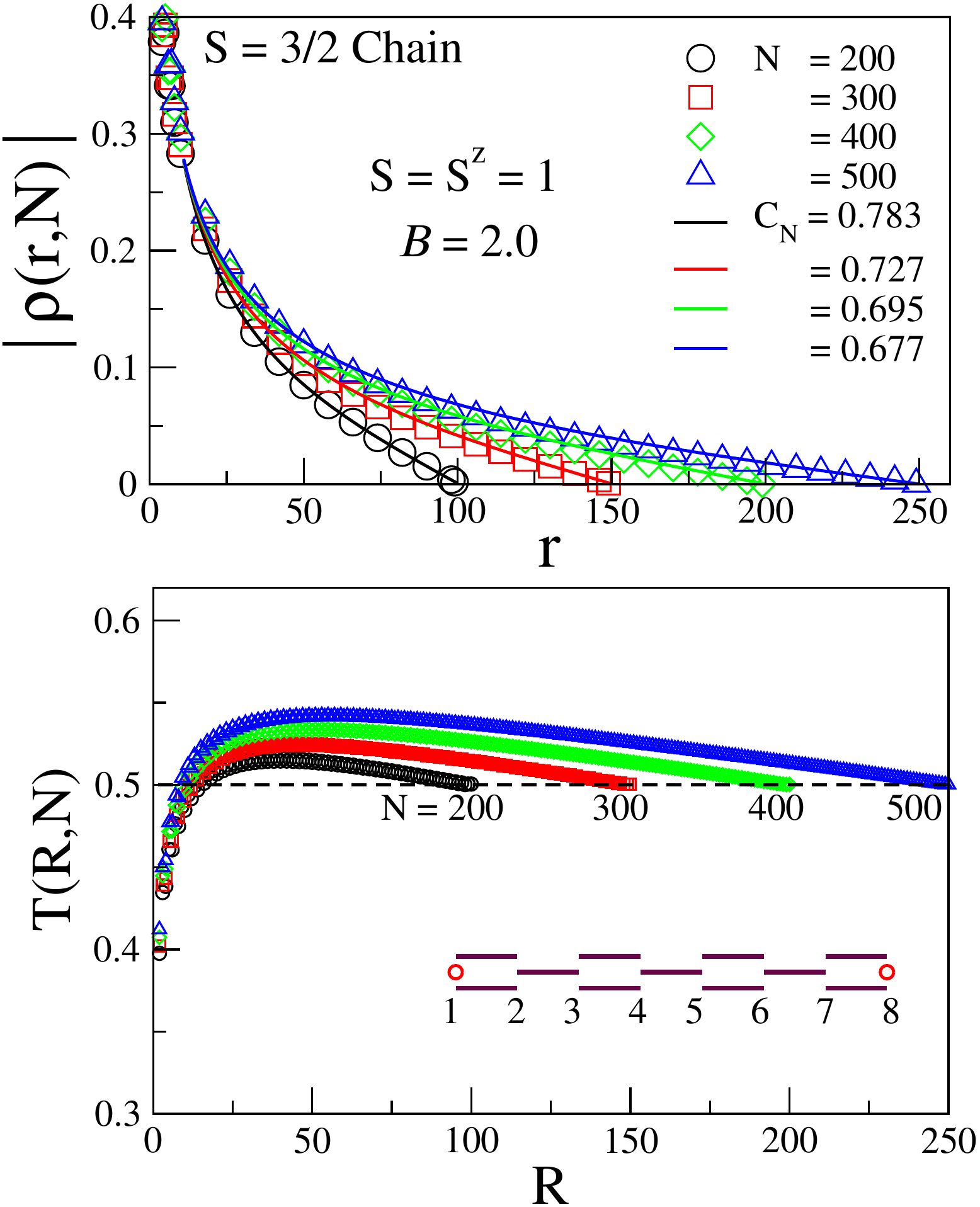}
\caption{\label{fig:s3by2-even}Upper panel: Open symbols are DMRG 
spin densities  $|\rho(r,N)|$ to the middle of $S = 3/2$ HAFs with 
even $N$ in Eq.~\ref{eq:ham}. The lines are fits based on 
Eq.~\ref{eq:spden-3by2} with $B = 2$ and the indicated scale factors 
$C_N$. Lower panel: Cumulative spin densities, Eq.~\ref{eq:spdsum}, 
up to site $R$; inset: VBS valence bond diagram for even $N$.}
\end{figure}

The BI-SDWs of even chains are triplets. The 
ratios $f(r,N)$ in Eq.~\ref{eq:fdef} are quite different for the 
$S = 3/2$ chain, either even or odd, and are not shown. The upper panel of 
Fig.~\ref{fig:s3by2-even} shows the magnitude of spin densities 
up to the middle of chains. The SDWs converge at small $r$ but are 
not localized in the $S = 3/2$ chain. The spin densities add up to 
$S^z = 1$ for even $N$. They decrease slowly and the sum over 
$|\rho(r,N)|$ diverges in the thermodynamic limit. The lines are 
fits that are discussed below. The lower panel of 
Fig.~\ref{fig:s3by2-even} shows the cumulative spin density to site 
$R$ that we define as
\begin{equation}
T(R,N) = \sum_{r=1}^R \rho(r,N) + \rho(R+1,N)/2.
\label{eq:spdsum}
\end{equation}
The total spin density is $S^z = 1 = T(N - 1,N) + \rho(1,N)/2$. 
$T(R,N)$ increases rapidly to 0.5 around $R_{1/2} \sim 15$, reaches 
a broad maximum that depends on system size and decreases as 
required by symmetry to $0.5$ in the middle of the chain. The 
VBS valence bond diagram in the inset has unpaired spins at each 
end that correspond~\cite{tkng94} to $s' = (S - 1/2)/2 = 1/2$. 
Each $S = 3/2$ site forms three singlet-paired spins to a neighbor. 
The middle and either the top or bottom line corresponds to the VBS 
diagram of the $S = 1$ chain with a localized spin at the ends. 
The remaining line with paired spins is a singlet valence bond diagram of 
the $S = 1/2$ chain. The slow variation of $T(R,N)$ in the middle and no 
net spin between $R_{1/2}$ and $N - R_{1/2}$ is consistent with 
singlet-paired spins.

\begin{figure}
\includegraphics[width=\columnwidth]{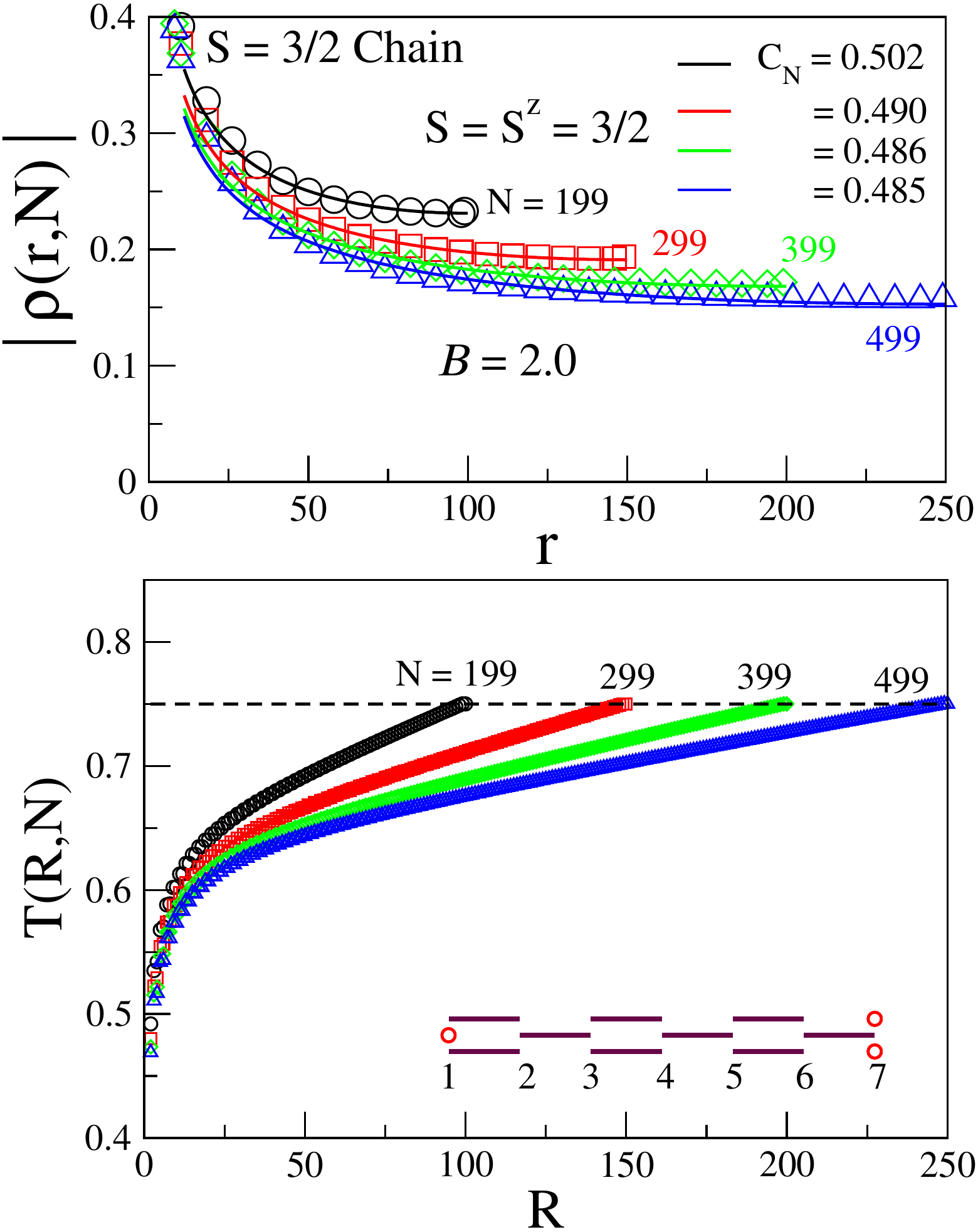}
\caption{\label{fig:s3by2-odd}Upper panel: Open symbols are
DMRG spin densities $|\rho(r,N)|$ to the middle of $S = 3/2$
HAFs with odd $N$ in Eq.~\ref{eq:ham}. The lines are fits
based on Eq.~\ref{eq:spden-3by2} with $B = 2$ and the indicated
scale factors $C_N$. Lower panel: Cumulative spin densities,
Eq.~\ref{eq:spdsum}, up to site $R$; inset: one of two 
equivalent VBS valence bond diagrams for odd $N$).}
\end{figure}

The GS of odd chains is a quartet, $S = 3/2$. Figure~\ref{fig:s3by2-odd}, 
upper panel, shows $|\rho(r,N)|$ to the middle of chains. The large 
amplitude of in-phase BI-SDWs in the middle decreases slowly with 
system size. The cumulative spin density $T(R,N)$ in the lower 
panel is again given by Eq.~\ref{eq:spden-3by2} except that the 
$r = (N + 1)/2$ spin density is shared equally between the two halves. 
The total is $S^z = 3/2$ for the entire chain, or 0.75 for the half 
chain. The rapid initial increase to $T(R_{1/2},N) = 0.5$ by 
$R_{1/2} \sim 15$ suggests a spin-1/2, as does the gradual increase 
to $0.75$ in the middle. The VBS valence bond diagram in the lower 
panel has three unpaired spins, two at one end, one at the other end; 
the diagram with reversed unpaired spins at the ends contributes 
equally by symmetry. The middle and either top or bottom line is again 
the $S = 1$ VBS diagram. The remaining line is an $S = 1/2$ 
valence bond diagram with an unpaired spin at either end. Although the diagram 
correctly has three unpaired spins, the DMRG spin densities clearly 
show one spin in the central region rather than at the ends.
  
The $S = 1/2$ HAF with odd $N$ does not support edge states. The 
spin density is delocalized over the entire chain.~\cite{soos83} 
Even more simply, a half-filled tight binding 
or H\"uckel band of $N = 2n + 1$ sites has spin density $1/(n + 1)$ 
at odd numbered sites and $\rho = 0$ at even numbered sites; in that 
case, $T(R,N)$ goes as $R/(n + 1)$ and immediately rationalizes the 
linear increase in Fig.~\ref{fig:s3by2-odd}, lower panel. We 
attribute the larger gap $|\Gamma_{3/2}(N)| > \Gamma_1(N)$ in 
Fig.~\ref{fig:gap-s3by2} and its weaker dependence of system size 
to enhanced coupling between the ends by the delocalized spin in 
the middle. 

The BI-SDW amplitude at the middle in the upper panel of 
Fig.~\ref{fig:s3by2-odd} decreases slightly faster than $N^{-1/2}$.
The size dependence of the amplitude suggests modeling 
the spin densities as
\begin{eqnarray}
\rho(r,N) &=& (-1)^{r-1} C_N \left( \left(\frac{\ln B r}{r}\right)^{1/2}\right.
\nonumber \\
& & \left.  -(-1)^N \left(\frac{\ln B(N+1-r)}{N+1-r}\right)^{1/2}\right).
\label{eq:spden-3by2}
\end{eqnarray}
The amplitude $C_N$ depends on system size because the SDWs are 
not localized. We took $|\rho(r,N)|$ with $B = 2$ and the 
indicated $C_N$ to generate the lines in the upper panels of 
Figs.~\ref{fig:s3by2-even}~and~\ref{fig:s3by2-odd}. The spin 
densities are adequately fit in the central region in either 
case. Deviations are seen for $r < 10$ when $N$ is even and for 
$r < 15$ when $N$ is odd.

To some extent, Eq.~\ref{eq:spden-3by2} can be understood in 
terms of the NL$\sigma$M. In the thermodynamic limit, the GS spin 
correlation functions $C(r)$ depend only on the separation $r$ 
between spins. The NL$\sigma$M result is~\cite{sorensen94}
\begin{equation}
C(r) \equiv \langle S_0^z S_r^z \rangle \propto (-1)^r r^{-1/2}
\exp (-r/\xi),
\label{eq:corl}
\end{equation}
for integer spin HAFs and $r \gg \xi$. Several 
authors~\cite{nomura89,white-huse-prb93,sorensen94} have remarked 
that DMRG results for $r^{1/2}|C(r)|$ are noticeably closer to 
exponential in $S = 1$ chains of 60 or 100 spins. Since converged 
$C(r)$ are limited to about $r < N/4$, such agreement is promising 
but not forced. White and Huse discuss~\cite{white-huse-prb93} the 
point explicitly and show (Fig. 4 of [\onlinecite{white-huse-prb93}]) 
that the ratio $|C(r)|$ to the NL$\sigma$M correlation function 
becomes constant at $r \sim 2 \xi \sim 12$. The first few sites 
where $C(r)$ can be computed  most accurately are inevitably excluded 
from direct comparison since the NL$\sigma$M describes a continuous 
rather than a discrete system. The $r^{-1/2}$ factor in $C(r)$ does 
not appear in the spin densities of integer $S$ chains,~\cite{sorensen94} 
whose exponential decrease with $r/\xi$ is shown in 
Figs.~\ref{fig:amp-s1}~and~\ref{fig:amp-s2}. The spin correlations 
of the $S = 1/2$ HAF go as $|C(r)| \propto (\ln r/ r_0)^{1/2}/r$ 
according to field theory~\cite{affleck89} and Monte Carlo 
calculations~\cite{sandvik2010} up to $N = 4096$ return $r_0 = 0.08$. 
But exact results for $C(r)$ in finite PBC systems~\cite{sandvik2010} 
still show significant deviations at $N = 32$. 

Hallberg et al.~\cite{karen96} applied the NL$\sigma$M and DMRG to 
the $S = 3/2$ chain and confirmed that it belongs to the same 
universality class as the $S = 1/2$ chain. They report 
$|C(r)| \propto (\ln B r)^{1/2}/r$ and estimate $B = 0.60$ from 
$r = 4$ to $25$ in a 60-spin chain. We find $B = 0.45$ in similar 
calculations for $N$ = 200. F\'ath et al.~\cite{fath2006} extrapolate 
to $B = 0.11$ for $\Gamma_1(N)$ in the thermodynamic limit. The 
differences are negligible in the context of spin densities. Then 
$r^{1/2} |C(r)|$ gives Eq.~\ref{eq:spden-3by2} when contributions 
from both ends are taken into account. DMRG results for $|\rho(r,N)|$ 
deviate from Eq.~\ref{eq:spden-3by2} near the ends of $S = 3/2$ 
chains and from Eq.~\ref{eq:spden-post} in integer $S$ chains. 
The choice of $B$ changes the fits at small $r$. Since small $r$ 
is not modeled quantitatively in either case and does not concern 
us here, we took $B = 2$ in Eq.~\ref{eq:spden-3by2} for the spin 
densities of $S = 3/2$ chains. 

Three effective spins are needed for the $S = 3/2$ spin densities 
when $N$ is odd, a spin $s'$ in the middle in addition to spins at 
the ends. The generalization of Eq.~\ref{eq:heff} to half integer 
$S$ and odd $N$ is
\begin{equation}
H_{eff}(N) = -J_1(N)\left(\vec{s'} \cdot \left(\vec{s'}_1 + \vec{s'}_N
 \right) \right) - J_2(N) \vec{s'}_1 \cdot \vec{s'}_N.
\label{eq:heff-odd}
\end{equation}
The eight microstates of $H_{eff}$ correspond to the GS quartet and 
two doublets. Both the total effective spin $S'$ and 
$S'_{1N} = s'_1 + s'_N$ are conserved, with $S' = 3/2$, $S'_{1N} = 1$ 
in the GS. The doublets have $S' = 1/2$ and $S'_{1N} = 0$ or 1. The 
spectrum is
\begin{eqnarray}
E_{eff}(S',S'_{1N}) &=& -\frac{J_1}{2} S'(S'+1) 
 + \frac{J_1 - J_2}{2} S'_{1N}(S'_{1N} + 1) \nonumber \\
& & \quad + \frac{3(J_1+2J_2)}{8}.
\label{eq:spectra}
\end{eqnarray}
The gap $\Gamma_{3/2} = -J_1/2 - J_2$ is to the doublet with singlet-paired 
spins at the ends; the gap to parallel spins is $-3J_1/2$. The 
effective exchanges in Eq.~\ref{eq:heff-odd} can be fit to DMRG 
results for the doublets with the lowest and second lowest energy in 
the $S^z = 1/2$ sector. We find $J_1(99) = 0.04214$, $J_2(99) = -0.00346$ 
and $J_1(199) = 0.01836$, $J_2(199) = -0.00176$.  Large $|\Gamma_{3/2}(N)|$
in Fig.~\ref{fig:gap-s3by2} for odd $N$ is due to $J_1(N)$ and coupling through 
the delocalized effective spin $s'$. The small effective exchange $J_2(N)$ 
is antiferromagnetic.

\begin{figure}[t]
\includegraphics[width=\columnwidth]{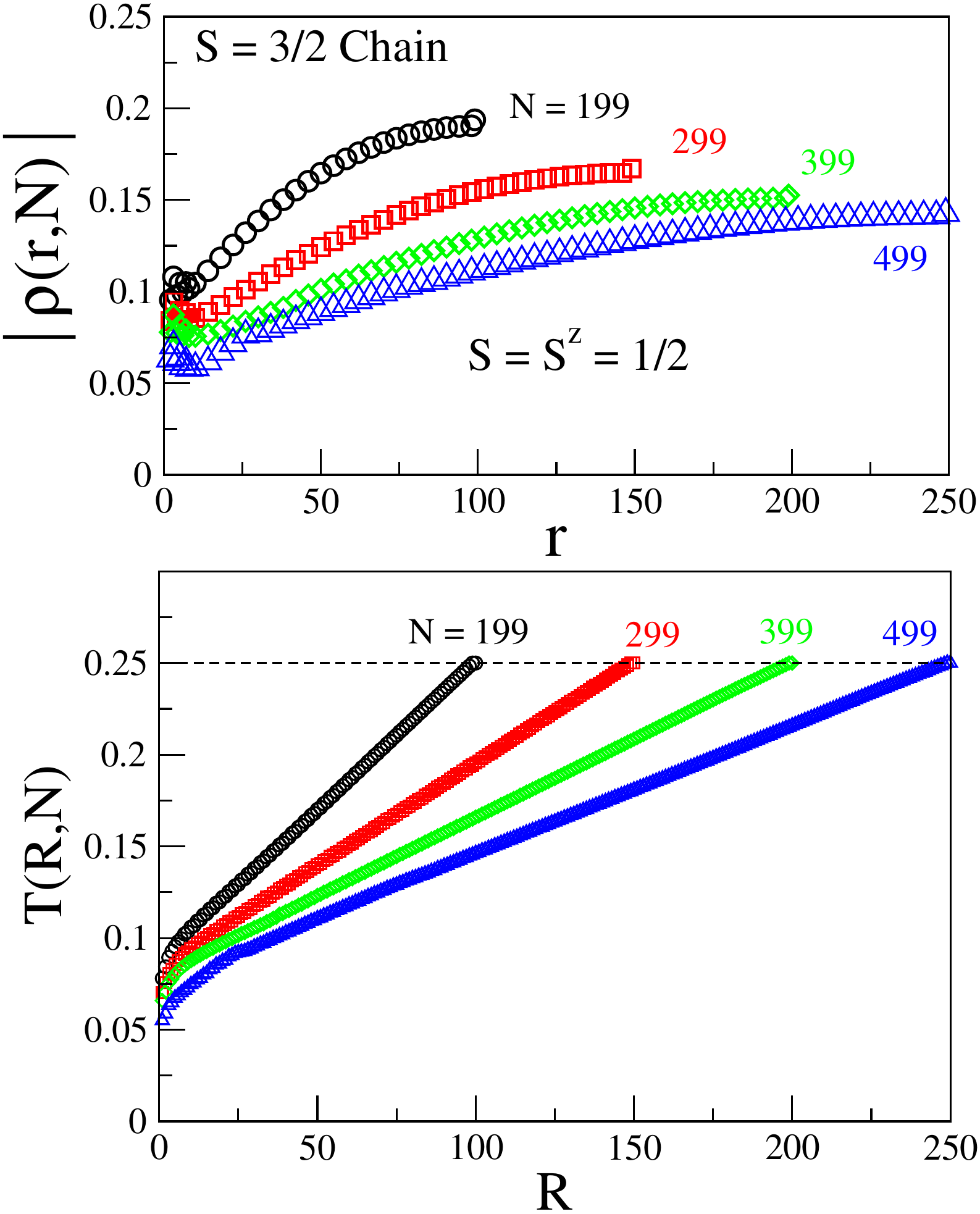}
\caption{\label{fig:s3by2-ex}Upper panel: DMRG spin densities 
$|\rho(r,N)|$ to the middle of $S = 3/2$ HAFs with odd $N$ 
in Eq.~\ref{eq:ham} for the lowest energy doublet state,
$S = S^z = 1/2$. Lower panel: Cumulative 
spin densities, Eq.~\ref{eq:spdsum}, up to site $R$.}
\end{figure}

To conclude the discussion of $S = 3/2$ chains, we recall that the GS for 
PBC and odd $N$ has $S_G = 1/2$. Since $J_{1N} = J$ is between sites in 
the same sublattice, the system is not bipartite, and the GS has a domain 
wall or topological soliton. The OBC system is bipartite. The doublet 
$S = S^z = 1/2$ with the lowest energy has positive spin densities at 
odd-numbered sites and negative spin densities at even-numbered sites, 
respectively, with singlet paired $s'_1$ and $s'_N$ in Eq.~\ref{eq:heff-odd}. 
Figure~\ref{fig:s3by2-ex} shows $|\rho(r,N)|$ for $S = S^z = 1/2$ to the middle 
of $S = 3/2$ chains in the upper panel and the cumulative spin density 
$T(R,N)$ in the lower panel. The magnitude of the spin density at the 
middle decreases roughly as $N^{-0.42}$. There are no boundary-induced 
edge states. The spin is delocalized as expected on general grounds and 
becomes the effective spin $s'$ in Eq.~\ref{eq:heff-odd}. By contrast, 
the spin densities are entirely associated with BI-SDWs in OBC systems 
with even $N$ or integer $S$ since singlet states have $\rho(r,N) = 0$ 
at all sites.

\section{\label{discussion}Discussion}
We have applied different DMRG algorithms to spin-$S$ HAFs, 
Eq.~\ref{eq:ham}, with even and odd number sites in order to 
obtain accurate edge states in chains of several hundred spins. 
The principal results are the energy gaps $\Gamma_S(N)$, Eq.~\ref{eq:gap}, 
and the spin densities $\rho(r,N)$, Eq.~\ref{eq:spden}, that 
are modeled as boundary-induced spin density waves (BI-SDWs) at 
both ends. For the $S = 1$ HAF, we reproduce and refine previous 
studies on even chains of 60 or 100 spins that exceed the 
correlation length $\xi = 6.048$ by an order of magnitude. We 
confirm that the gap goes as $(-1)^N J_e \exp(-N/\xi)$ in chains 
with odd $N$. Two parameters, $\xi$ and the SDW amplitude, account 
quantitatively for $\Gamma_1(N)$ and $\rho(r,N)$ for chains from 
$N = 35$ to at least 120. The BI-SDWs are in phase for odd $N$, out 
of phase for even $N$. 

The smaller Haldane gap of the $S = 2$ HAF or the gapless $S = 3/2$ 
HAF requires substantially longer chains, here up to 500 spins, 
whose edge states have previously been studied in shorter chains 
$N < 200$. The spin densities of $S = 2$ HAFs beteen $N = 199$ and 
502 are modeled by BI-SDWs with correlation length $\xi = 49.0$ and 
amplitude $A = 0.90$. There are now two gaps, $\Gamma_1(N)$ and 
$\Gamma_2(N)$, that decrease exponentially as $r / \xi$ up to the 
$N \sim 220$ limit of our numerical accuracy. The gap ratio is 
$\Gamma_2(N)/\Gamma_1(N) = 3.45$. The gap $\Gamma_1(N)$ of the 
$S = 3/2$ HAF with even $N$ decreases faster than $1/N$, roughly as 
$N^{-1.42}$ or as $1/\ln(0.11 N)$. The gap $\Gamma_{3/2}(N)$ for odd 
$N$ has larger amplitude and weaker size dependence. The BI-SDWs of 
the $S = 3/2$ chain have maximum spin density at the ends but are not 
localized. The $S = 3/2$ spin densities in chains of more than 100 
spins have not been previously reported to the best of our 
knowledge. The $S = 3/2$ ground state for odd $N$ can be modeled as a 
spin-1/2 at each end and a spin-$1/2$ in between. 
 
DMRG calculations can be performed on longer chains of $N \sim 1000$ 
and/or larger $S$. But the condition $N > 5 \xi$ for integer $S$ is 
increasingly difficult to satisfy for small Haldane gaps $\Delta(S)$ 
whose rapid decrease has been reported~\cite{nakano2009} to $S = 5$. 
Moreover, the gaps will require extraordinary accuracy since, as 
shown in Table~\ref{tab:gaps}, $\Gamma(N) < 10^{-3}$ is reached at 
$N = 200$ for $S = 2$ or at $N = 400$ for $S = 3/2$. Spin densities 
are more promising probes of long chains in terms of the $S^z > 0$ 
sectors of $N$ and $N - 1$ spins. But the required system size for 
half integer $S$ is poorly known and may not have been reached in 
the present work. 

The nonlinear sigma model (NL$\sigma$M) and valence bond solid (VBS) 
have been applied to spin chains, primarily to the $S = 1$ HAF in the 
thermodynamic limit. Machens et al.~\cite{machens2013} summarize and 
critically evaluate both the NL$\sigma$M and VBS in connection with 
short chains of less than 20 spins. In partial disagreement with earlier 
works, they find that the effective coupling between edge states in 
Eq.~\ref{eq:heff} in short chains is influenced by the comparably small 
finite-size gaps of bulk excitations. We have characterized long chains whose 
prior modeling has mainly been for $S = 1$. 

Accurate DMRG results for 
$S = 2$ or $S = 3/2$ HAFs are a prerequisite for comparisons, mainly 
via $H_{eff}$ in Eq.~\ref{eq:heff}, with either the NL$\sigma$M or VBS. 
Good agreement in $S = 1$ chains carries over to some extent to $S = 2$ 
chains and less so to $S = 3/2$ chains. Spin densities to $N = 500$ 
yield $\xi = 49.0$ for the correlation length of the $S = 2$ chain. 
The gaps in shorter chains return the same $\xi$, but the ratio 
$\Gamma_2(N)/\Gamma_1(N)$ is 3.45 instead of 3. The deviation is real. 
The 3/2 chain does not follow the~\cite{lou2002} $(-1)^NJ_e(N)$ pattern 
of $H_{eff}$ in Eq.~\ref{eq:heff}. The BI-DWIs are not localized. Two 
effective spins $s' = 1/2$ at the ends account for $\Gamma_1(N)$ when 
$N$ is even. A third $s' = 1/2$ in the middle leads to $\Gamma_{3/2}(N)$ 
and $H_{eff}$ in Eq.~\ref{eq:heff-odd} for odd $N$. 

In other ways, however, comparisons are simply not possible. Since 
field theory starts with a continuous system rather than a discrete 
chain, the ends can be distinguished from the bulk but not sites at 
a finite distance from the ends. Similarly, VBS deals with special 
Hamiltonians,~\cite{affleck88,machens2013,schollwock96,totsuka95} 
that contain, in addition to Eq.~\ref{eq:ham}, terms that go as 
$B_p({\mathbf S}_r \cdot {\mathbf S}_{r+1})^p$ with $2 \le p \le 2S$ 
and coefficients $B_p$. Exact GS are obtained in the thermodynamic 
limit of these models. The relevant valence bond diagrams 
have paired spins, as shown, except at the first and last sites.
Either the NL$\sigma$M or VBS correctly places localized states or 
unpaired spins for integer $S$, but neither describes the BI-SDWs 
found in DMRG calculations spin-$S$ HAFs. The BI-SDWs are not localized 
in half integer $S$ chains and have different effective coupling 
between ends. Direct solution of Eq.~\ref{eq:ham} for $S \ge 1$ 
chains inevitably leads to edge states whose features are blurred or 
lost in the NL$\sigma$M or VBS. Comparisons may well be limited to 
effective spins and exchange at the ends. 

The occurrence of edge states in HAFs with $S \ge 1$ follows directly 
from Eq.~\ref{eq:ham}, as shown in the Introduction. PBC systems 
with $J_{1N} = J$ have $S_G = 0$ except for half integer $S$ and odd 
$N$, when $S_G = 1/2$. OBC systems with $J_{1N} = 0$ have $S_G = 0$ 
for even $N$ and $S_G = S$ for odd $N$. The energy per site in the 
thermodynamic limit cannot depend on boundary conditions for 
short-range interactions. Different $S_G$ under OBC and PBC implies 
edge states, or BI-SDWs, in HAF with $S \ge 1$ and gaps $\Gamma_S(N)$ 
relative to $S_G = 0$ for even $N$ or 
integer S or to $S_G = 1/2$ for odd $N$ and half integer S. 
The size dependence and interpretation of gaps or spin densities are 
standard for integer $S$. The spin densities and gaps of the $S = 3/2$ 
chain lead to different BI-SDWs for even and odd $N$. The NL$\sigma$M 
or VBS provides useful guidance for quantitative modeling 
of BI-SDWs obtained by DMRG for HAFs with $S \ge 1$.

\begin{acknowledgments}
We thank S. Ramasesha and D. Huse for discussions and the NSF for 
partial support of this work through the Princeton MRSEC 
(DMR-0819860). MK thanks DST for a Ramanujan Fellowship SR/S2/RJN-69/2012
and DST for funding computation facility through SNB/MK/14-15/137.
\end{acknowledgments}

\bibliography{edge-ref}

\end{document}